\documentstyle[prl,aps,multicol,epsf]{revtex}
\tighten
\begin{document}
\draft
\title{Mesoscopic fluctuations of tunneling through double quantum dots}
\author{A. Kaminski and L.I. Glazman}
\address{Theoretical Physics Institute, University
of Minnesota, Minneapolis, MN 55455}
\maketitle
\begin{abstract}                  
We study fluctuations of conductance of two connected in series
dots in the Coulomb blockade regime. The pattern of the fluctuations
turns out to be extremely sensitive to a magnetic field.   These
conductance fluctuations also provide information about the
mesoscopic fluctuations of charge of a partially-opened single quantum dot,
which are hard to measure directly.
\end{abstract}
\begin{multicols}{2}
Charge of a mesoscopic conductor attached to leads by high-resistance
junctions is quantized. This quantization  (Coulomb blockade) is progressively
smeared with the reduction of the junction resistance to the level $\sim
h/e^2$. Most directly, the smearing can be studied by a measurement of
charge or differential capacitance of a quantum dot  connected to a particle
reservoir by a tunable junction of  conductance
$G_i$. This proved to be a hard experiment  however, performed only very
recently\cite{Zhitenev}.

 Spatial quantization of the electron motion in a dot leads
to mesoscopic fluctuations of its transport
\cite{JalabertEtal92}  and thermodynamic
\cite{Buttiker,KaminskiEtal98} characteristics in the
Coulomb blockade regime.  Fluctuations of the conductance in
a single dot in both peaks and valleys of the Coulomb
blockade are of  the order of the average conductance, and
were studied experimentally in Refs.
\cite{Chang}.  Fluctuations of the ground
state energy of a partially  opened dot result in a random
contribution to its charge and differential capacitance.
However these fluctuations are small\cite{KaminskiEtal98},
therefore hard to observe  directly. The accuracy of the
experiments \cite{Zhitenev}, where the average
differential capacitance was studied, is insufficient for
quantitative study of the fluctuations.

\begin{figure}
\epsfxsize=8.5cm
\centerline{\epsfbox{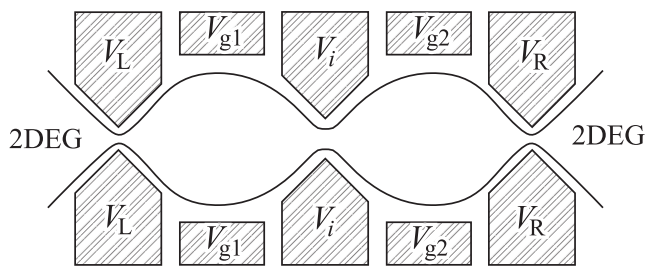}}
\caption{\label{fig:Double dot} Schematic view of the
double-dot system. The dots are formed by applying a
negative voltage to the gates. 
The conductance $G_L$ and $G_R$ of the channels connecting the double-dot to
the electron reservoirs is controlled by 
$V_L$ and $V_R$; voltage $V_i$ controls the
conductance $G_i$ of the interdot channel. Voltages
$V_{g1}$ and $V_{g2}$ are used to change the number of
electrons in the dots.}
\end{figure}

 In this paper we demonstrate that a number of features of
the mesoscopic fluctuations in the Coulomb blockade regime
is much easier to observe in a double-dot geometry,  shown
in Fig.~\ref{fig:Double dot}. The dots are statistically
identical, with average level spacing $\Delta$, and charging
energy $E_C$. The temperature $T$ is much smaller than
$\Delta$ but exceeds the level widths associated with
electron escape to the leads. Conductances of the three
junctions satisfy relation $e^2/h \gtrsim G_i\gg G_L,\,G_R$.
The dependence of the conductance $G$ of the system on
$V_{g1}$ and $V_{g2}$ in this case reveals sharp twin
peaks\cite{LivermoreEtal96}.

We calculate the  variance of the peak splitting and show
that it gives a  direct measure of the random component of
the ground state energy and charge of a partially opened
dot.  The positions of the peak doublets with respect to each
other also fluctuate.  In the absence of the fluctuations,
the peaks would form a regular pattern (see Fig.~\ref{fig:2D
peaks}$a$). Random displacements of peak doublets manifest
themselves most prominently in strong fluctuations of the
conductance measured along the  line $V_{g1}=V_{g2}\equiv
V_g$ on the plane $(V_{g1},V_{g2})$. These fluctuations are
unusually sensitive to a magnetic field. The
corresponding correlation field is parametrically smaller
than the field leading to crossover from orthogonal to
unitary ensemble.

\narrowtext

\begin{figure}
\epsfxsize=8.5cm
\centerline{\epsfbox{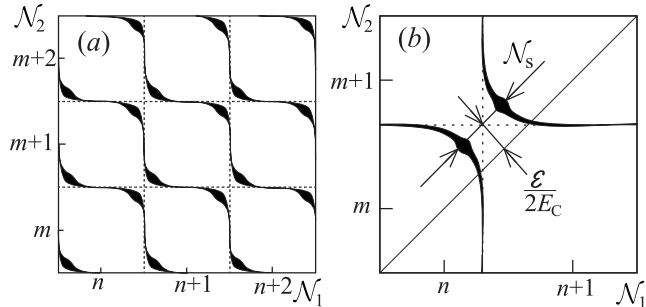}}                      
\caption{\label{fig:2D peaks}
($a$) The map of the conductance of the double-dot system (after
\protect\cite{LivermoreEtal96}). The conductance ridges are shown with solid
lines; the width of a line increases with the ridge height. 
($b$) The structure of a peak
doublet, displaced from the line ${\cal N}_{1}={\cal N}_{2}$ due to the
ground state energy fluctuations.}
\end{figure}

{\em Qualitative consideration and the main results ---} The
dependence of  conductance $G$ of the double-dot system on
the gate voltages $(V_{g1},V_{g2})$ forms a
three-dimensional landscape. The ridges on this landscape,
shown by solid lines on the map Fig.~\ref{fig:2D peaks}$a$,
correspond to the activationless hopping of an electron on
and off the double dot. When $G_i\to 0$, the ridges form a
rectangular pattern (dashed lines on  Fig.~\ref{fig:2D
peaks}$a$). The conductance has
peaks when  both dots are at resonance (crossings of the
dashed lines). At these special values of the gate voltages,
the four states with $n$ or $n+1$ electrons in the left dot
and $m$ or $m+1$ electrons in the right dot have the same
energy.  Finite inter-dot tunneling ($G_i \ne 0$) removes this degeneracy.
The two states with the total number of electrons $n+m+1$
hybridize, lowering the energy of the  ground state by
$U$, with $\langle U\rangle =(\ln  2/\pi^2)(hG_i/e^2)E_C$\cite{MatveevEtal96}.
This  results in anticrossing of the conductance ridges and
formation of twin peak structures shown in Fig.~\ref{fig:2D
peaks}. The peak splitting (see Fig.~\ref{fig:2D peaks}$b$) 
in dimensionless units ${\cal N}=C_gV_g/e$ is
\begin{equation}
{\cal N}_s=\frac{2CU}{e^2}\,,\,\mbox{with}\;
\langle {\cal N}_s\rangle =\frac{\ln 2}{\pi^2}\frac{hG_i}{e^2}\;,
\label{Vs}
\end{equation} 
where $C_g$ and $C$ are the
capacitance to the gate and the total capacitance of a
single dot respectively, and $\langle...\rangle$
denotes the ensemble averaging.

\narrowtext

In random dots, the inter-dot tunneling amplitudes
fluctuate, giving rise to a random contribution $\delta U$
to the energy $U$. Let us estimate $\delta U$. The tunneling
correction to the ground state energy results from the
virtual electron transitions between the dots. The number of
one-electron states contributing to the correction is $\sim
E_C/\Delta$\cite{KaminskiEtal98}. The contribution of each
state fluctuates, with the magnitude of the order of its
average, because of the fluctuations of the wave functions at
the point of the inter-dot contact. As the result, 
\begin{displaymath}
\langle\delta U^2\rangle\sim 
\frac{\Delta}{E_C}\langle U\rangle^2\simeq
\left(\frac{hG_i}{e^2}\right)^2E_C\Delta\;.
\end{displaymath}
However, in a single-dot geometry, this quantity is
accessible only through a difficult measurement of the
differential capacitance. In the structure of
Fig.~\ref{fig:Double dot},  the mesoscopic fluctuations of the ground state
energy of a partially opened quantum
dot lead to the fluctuations of
peak splitting,
\begin{equation}
\frac{\langle\delta {\cal N}_s^2\rangle} {\langle {\cal N}_s\rangle^2}\approx \frac{2}{\beta}
\frac{\Delta}{E_C}\;.
\label{deltaN}
\end{equation}
 Here $\beta=1$ or $2$ for the orthogonal or
unitary ensemble respectively. 
The numerical coefficient in
Eq.~(\ref{deltaN}) is obtained in {\em Quantitative
description}. Thus the double-dot layout allows one to study
the fluctuations of the ground state energy by means of
relatively simple transport measurements.

We consider also the random displacements of peak doublets with respect to each 
other.  In an ideal structure, peak doublets form a square lattice, and the line 
${\cal N}_{1}={\cal N}_{2}$  traverses exactly through the apexes of the split 
conductance  peaks. In random dots, however, the energies $E(n,m)$ fluctuate, so 
these peak doublets are randomly displaced away from the line ${\cal 
N}_{1}={\cal N}_{2}$ (Fig.~\ref{fig:2D  peaks}$b$).  We will show that typically 
the displacement ${\cal E}$ of a peak exceeds considerably its width. It leads 
to a wide distribution of the heights $\tilde{G}_{max}$ of maxima in the  
conductance measured along this line, $\tilde{G}({\cal N})\equiv G({\cal 
N},{\cal N})$. (Note the difference in definitions of $\tilde{G}_{\it max}$ and the
height of a peak in $G({\cal N}_1, {\cal N}_2)$).

The displacement ${\cal E}$ is a sum of two independent terms. One of them, 
${\cal E}_1\propto\delta U$, stems from the fluctuations in the interdot 
interaction. The second one, ${\cal E}_2\sim\Delta$, is due to the 
randomness of one-electron spectra in the dots \cite{Dyson62}.  The peak width 
in the direction perpendicular to the line ${\cal N}_1={\cal N}_2$ is 
determined by the hybridization between the discrete levels mediating the 
tunneling through the double-dot, and is of the order of 
$\sqrt{hG_i/e^2}\Delta$, which is much smaller than $\sqrt{\langle{\cal 
E}_1^2\rangle +\langle{\cal E}_2^2\rangle}$. The ratio of the two latter 
parameters yields the probability for the line ${\cal N}_1={\cal N}_2$ to 
``meet'' a peak, and thus describes the fluctuations of $\tilde{G}_{max}$,  
\begin{displaymath}
\frac{\langle \tilde{G}_{max}^2\rangle}{\langle
\tilde{G}_{max}\rangle^2}\sim
\frac{\sqrt{\langle{\cal E}^2\rangle}}{\sqrt{hG_i/e^2}\Delta}\gg 1\;.
\end{displaymath} 
The exact result obtained in {\em Quantitative
discussion} differs from the above estimate only by numerical factors:
\begin{equation}
\frac{\langle \tilde{G}_{max}^2\rangle}{\langle
\tilde{G}_{max}\rangle^2}\simeq A\sqrt{
0.25\frac{hG_i}{2e^2}\frac{E_C}{\Delta}+
\frac{2e^2}{hG_i}\ln\frac{E_C}{\Delta} }\;.
\label{Ghoba}
\end{equation}                           
Here $A\approx 0.45$ and $A \approx 0.23$ for the
orthogonal and unitary ensembles respectively.
We see that the variance of $\tilde{G}_{\it max}$ is parametrically
larger than $\langle \tilde{G}_{max} \rangle$, in all the
considered domain $G_i\lesssim e^2/h$.

The characteristic magnetic field altering the pattern of fluctuations in 
$\tilde{G}({\cal N})$ also can be obtained from the above qualitative 
consideration. The applied magnetic field shifts the two discrete levels 
mediating the conductance \cite{SimonsAltshuler93}.   A shift exceeding the peak 
width leads to a considerable suppression of a maximum conductance  
$\tilde{G}_{\it max}$. Such a shift can be produced by a small variation of the 
magnetic field,
\begin{equation}
B_c=\frac{\Phi_0}{S}\left[\frac{hG_i}{4\pi^2e^2}
\frac{\Delta}{4\pi E_T}\right]^{1/2} \;.
\label{Bc}
\end{equation}
Here $S$ and $E_T$ are
the area and the Thouless energy \cite{Thouless} of a single dot, $\Phi_0$
is the flux quantum. Field $B_c$ is much smaller than the
characteristic magnetic field $B_\Delta=(\Phi_0/S)\sqrt{\Delta/4\pi
E_T}$ changing the electron states in a single dot\cite{FalkoEfetov94}.
Therefore $B_c$ does not lead to the crossover from the orthogonal to unitary
ensemble.

 {\em Quantitative description ---}  To calculate the
conductance of the double-dot system, we use the Hamiltonian
in the following form
\begin{eqnarray}
\hat{H}&=&\hat{H}_d+\hat{H}_C+\hat{H}_{\it lead}+\hat{H}_t\;,\label{H}\\
\hat{H}_d&=&\sum_k \xi^{\phantom{\dagger}}_k a_k^\dagger
a_k^{\phantom{\dagger}}+ \sum_l \xi^{\phantom{\dagger}}_l a_l^{\dagger}
a^{\phantom{\dagger}}_l
\label{Hd}\;,\\
\hat{H}_t&=&\sum_{kp}\left[t_{kp}^{\phantom{\dagger}}a^\dagger_k
c^{\phantom{\dagger}}_{p}+h.c.\right]
+\sum_{lq}\left[t_{lq}^{\phantom{\dagger}}a^\dagger_l
c^{\phantom{\dagger}}_q +h.c.\right]
\nonumber\\
&+&\sum_{kl}\left[t_{kl}^{\phantom{\dagger}}a^\dagger_k
a^{\phantom{\dagger}}_{l} + h.c.\right]\;,\label{Ht}\\
\hat{H}_C&=&E_C \left[\left(\hat{n}_1-{\cal N}_{1}\right)^2
+\left(\hat{n}_2-{\cal N}_2\right)^2\right] \;.
\label{HC}
\end{eqnarray}
The subscripts $k$ and $l$, and $p$ and $q$ denote the
states in the left and right dot, and left and right lead
respectively;
 $\xi_{k,l}$ are the one-electron energy
spectra of the dots, $n_{L,R}$ are the numbers of electrons
in the left and right dot. The dimensionless gate voltages are
${\cal N}_i=C_gV_{gi}/e$, and $E_C=e^2/2C$ is the charging energy
of each dot; $C_g$ and $C$ are the
capacitance to the gate and the total capacitance of a
single dot. The effects of the interdot capacitance are
neglected. We also neglect the capacitance
of a dot to the gate of the other dot, which just skews the
map of Fig.~\ref{fig:2D peaks}. The Hamiltonian of isolated
leads is denoted $\hat{H}_{\it lead}$. 
The conductance of the
interdot junction is related to the matrix elements $t_{kl}$
by $G_i=(2e^2/h)\langle |t_{kl}|^2\rangle/(2\pi\Delta)^2$.
These matrix elements are proportional to 
the electron wave functions $\psi_{k,l}$
in the left and right dot at the point of
the interdot contact ${\bf r}_i$:
\begin{equation}
t_{kl}\propto \psi_k^*({\bf r}_i)\psi_l({\bf r}_i)\;.
\label{tkl}
\end{equation}
The matrix elements $t_{kp}$ and
$t_{lq}$ obey similar relations. 
Using the Hamiltonian (\ref{H})--(\ref{HC}) we find the
conductance $G$ of the system. The randomness of the one-electron spectra and
wave functions results in the fluctuations of the heights and positions of the
conductance peaks.

First we consider the mesoscopic fluctuations of the conductance peak
splitting ${\cal N}_s$, which is related to $U$ by Eq.~(\ref{Vs}).
To find $U$ we have to calculate
the  corrections to the energy of the four participating  states near the point of
their degeneracy, since
$U=[E^{(2)}(n+1,m)+E^{(2)}(n,m+1)]-[E^{(2)}(n,m)+E^{(2)}(n+1,m+1)]$. 
  As long as the
interdot  conductance $G_i$ is small, the corrections can be calculated by
the  second-order perturbation theory in $t_{kl}$ \cite{MatveevEtal96}.   As an 
example we give here the correction only to $E(n,m)$:
\begin{equation}
 E^{(2)}(n,m)=\sum_k \sum_l \frac{|t_{kl}^2|}{|\xi_k-\xi_l|+2E_C}
\theta(-\xi_k\xi_l).
\label{example}
\end{equation}
The Heaviside function here appears as a combination of the Fermi filling
factors at zero temperature.   The mesoscopic fluctuations in $U$ occur mainly
due to the fluctuations of the one-electron wave functions $\psi_{k,l}({\bf
r}_i)$, see Eq.~(\ref{tkl}) . The fluctuations of $\xi_{k,l}$ yield a smaller
contribution\cite{KaminskiEtal98}. Using the Porter-Thomas statistics for the
wave functions $\psi_{k,l}$ we arrive at
\begin{equation}
\langle \delta U^2\rangle=
\frac{2}{\beta} \left[\frac{1}{4\pi^2}-\left(\frac{\ln
2}{\pi^2}\right)^2\right] \left(\frac{hG_i}{2e^2}\right)^2
E_C\Delta\;,
\label{deltaU}
\end{equation}                           
which leads to Eq.~(\ref{deltaN}).
The corrections (\ref{example}) to the ground state energy are given by a sum
over the one-electron levels with energies $\lesssim E_C$.  Since there are
many statistically independent terms in this sum, the distribution of the
fluctuations $\delta U$ is Gaussian.

Now we will consider the fluctuations of conductance $\tilde{G}({\cal 
N})\equiv G({\cal N},{\cal N})$. The height of its maxima is a random 
quantity, depending not only on the one-electron wave functions, but also on the
random  displacement ${\cal E}$ of a peak from the  line ${\cal N}_{1}={\cal
N}_{2}$. If ${\cal E}<\Delta$, the electron transport occurs by  resonant
tunneling via two states, $k_0$, $l_0$. The height of the  conductance
maximum is then given by
\begin{equation}
\tilde{G}_{max}=\frac{2e^2}{h}
\frac{\Gamma_{k}\Gamma_{l}\displaystyle\frac{2\sqrt{2}|t_{kl}|^2} {\sqrt{{\cal
E}^2+2|t_{kl}|^2}} } {\Gamma_{k}+\Gamma_{l}
+\displaystyle\frac{(\Gamma_{k}-\Gamma_{l}){\cal E}} {\sqrt{{\cal
E}^2+2|t_{kl}|^2}} } \frac{\pi}{2T} \;,
\label{G1}
\end{equation}                          
with $k=k_0$, $l=l_0$. Here $\Gamma_k\propto|\psi_k({\bf
r}_L)|^2$ and $\Gamma_l\propto|\psi_l({\bf r}_R)|^2$ are the
widths of the states $k$ and $l$ with respect to the
electron tunneling into the adjacent lead; ${\bf r}_L$,
${\bf r}_R$ are the points of the dot-lead contacts. At
${\cal E}>\Delta$ there can be no inter-dot resonance on the  line ${\cal
N}_{1}={\cal N}_{2}$, and
the conduction process inevitably involves
cotunneling\cite{AverinNazarov90} through one of the dots.
At low temperature $T\lesssim \sqrt{E_C\Delta}$ only the
elastic cotunneling is possible. The resulting
conductance $\tilde{G}_{max}\sim (\Delta/{\cal E})G_i G_{L,R}$ yields
parametrically small contributions to the ensemble averages.
To calculate $\langle\tilde{G}_{\it max}\rangle$ and
$\langle\tilde{G}_{\it max}^2\rangle$, one needs to use Eq.~(\ref{G1})
and average over all the random quantities entering in it. The statistical
properties of
$\psi_{k,l}$ are well-known, and now we discuss those of ${\cal E}$.

One source of fluctuations of ${\cal E}$ is
the randomness of the interdot tunneling matrix elements,
which leads to random shifts of energies of electron states, see Eqs.~(\ref{tkl}),
(\ref{example}). This contribution yields:
\begin{equation}
\langle {\cal E}_1^2\rangle=\langle \delta U^2\rangle/2\;.
\label{E1}
\end{equation}
Another random contribution, ${\cal E}_2$,
comes from the fluctuations of the positions of the
individual levels in the dots. Its averaging is not quite trivial. In a real
experiment, the ensemble averaging $\langle\dots\rangle$ of peak characteristics
is replaced by averaging over a sequence of conductance peaks, and/or over a
range of magnetic field $B$. To study fluctuations within the orthogonal
ensemble, one must avoid using $B$ for acquiring the statistics of fluctuations.
Averaging over the sequence of maxima along the line ${\cal N}_1={\cal N}_2$ is not
actually a well-defined procedure, because ${\cal E}_2$ is not a
self-averaging quantity. Its average diverges as the number $N$ of
conductance peaks taken for averaging, increases, $\langle
[{\cal E}_2(N)]^2\rangle\propto\ln N$ \cite{Dyson62}. In this work we
refer to the following scheme of averaging: We choose an arbitrary
conductance doublet with some coordinates $({\cal N}_{1}^{(0)},{\cal
N}_{2}^{(0)})$ and then study $N$ consequent conductance maxima along the
line $({\cal N}_{1}^{(0)}+{\cal N},{\cal N}_{2}^{(0)}+{\cal N})$.
The number of maxima $N$ should be large enough to exceed the ``correlation
length''  the ground state energy. In a partially open quantum dot, the
characteristic correlation length of the fluctuations of the ground state
energy at different gate voltages is $\sim N_{\it corr}=E_C/\Delta$
\cite{KaminskiEtal98}. Thus, one must average over $N\simeq N_{\it corr}$
peaks which gives 
\begin{equation}
\langle{\cal E}_2^2\rangle=\frac{4\Delta^2}{\beta\pi^2}\ln
\left(\frac{E_C}{\Delta}\right),
\label{E2}
\end{equation}
The distribution of ${\cal E}_2$ is
Gaussian, except for the tails that are unimportant for our
purposes.

The distribution of the energy mismatch ${\cal E}$ is Gaussian;
its variance is $\langle{\cal E}^2\rangle=
\langle{\cal E}_1^2\rangle+\langle {\cal E}_2^2\rangle$, since the two contributions
are independent. The value of ${\cal E}$ is determined by a large number of
one-electron states.  Therefore one can neglect the correlations between 
${\cal E}$ and the one-electron wave functions of two states $k_0$ and $l_0$,
involved in Eq. (\ref{G1}). Averaging of Eq. (\ref{G1}) yields:
\begin{eqnarray}
\langle \tilde{G}_{max}\rangle&=& A_1 G_o\sqrt{\frac{hG_i}{2e^2}}
\frac{\Delta}{\sqrt{\langle{\cal E}^2\rangle}} \frac{\Delta}{4T}\;,
\label{Gav}\\
\langle
\tilde{G}_{max}^2\rangle &=&A_2 G_o^2 \sqrt{\frac{hG_i}{2e^2}} \frac{\Delta}
{\sqrt{\langle{\cal E}^2\rangle}} \left(\frac{\Delta}{4T}\right)^2\;.
\label{G2av}
\end{eqnarray}
The conductances of the dot-lead contacts for simplicity
here are taken equal: $G_L=G_R=G_o$. The numerical factors
are ensemble-dependent; $A_1=1/\sqrt{2}\pi^2$ (GOE) or
$A_1=\pi^2/64\sqrt{2}$ (GUE), and
\begin{displaymath}
A_2=\frac{3\pi-8}{\sqrt{\pi}16\pi^2} \mbox{ (GOE) or }
A_2=\frac{(32-9\pi)\sqrt{\pi}}{1024} \mbox{ (GUE).}
\end{displaymath}
Substitution of $\langle{\cal E}^2\rangle$ [Eqs. (\ref{E1}), (\ref{E2})]
into Eqs. (\ref{Gav}), (\ref{G2av}) gives Eq.~(\ref{Ghoba}).

The fluctuations of $\tilde{G}_{\it max}$ are very sensitive to the magnetic
field because the leading contribution to $\langle \tilde{G}^2_{\it
max}\rangle$ comes from the peaks, whose narrow resonant region is
traversed by the line ${\cal N}_{1}={\cal N}_{2}$. A variation of the magnetic
field by $B_c$ [Eq.~(\ref{Bc})] would shift \cite{SimonsAltshuler93} these peaks
away from the line by $\sim\!\sqrt{hG_i/e^2}\Delta$. 
This leads  to a significant decrease of the height of the
corresponding maximum in $\tilde{G}({\cal N})$. The one-electron wave
functions are not changed by such a small 
variation of the magnetic field. Therefore the
correlation function of the peak conductance at different magnetic
fields, $\langle
\tilde{G}_{peak}(B_1)\tilde{G}_{max}(B_2)\rangle$, is controlled by the 
parametric correlations of ${\cal E}_1$ with itself
at  two different fields. If $|B_1-B_2|\ll
B_\Delta\equiv(\Phi_0/S)\sqrt{\Delta/4\pi E_T}$, the distribution function of
the field-induced variation $\varepsilon(B_1-B_2)$ of the energy ${\cal E}$
is\cite{SimonsAltshuler93}
\begin{eqnarray}
&&P(\varepsilon)=\frac{1}{\sqrt{2\pi}\alpha}
\exp\left[-\frac{\varepsilon^2}{2\alpha^2}\right]\,,\;\label{PB}\\
&&\alpha=(\Delta/\sqrt{2})\left(|B_1-B_2|/B_\Delta\right)\;.\nonumber
\end{eqnarray}
Using Eqs.~(\ref{G1})-(\ref{E2}), (\ref{PB}), we obtain the correlation function $\langle
\tilde{G}_{max}(B_1)\tilde{G}_{max}(B_2)\rangle$.                       
Its full-width-half-maximum is $\approx 6B_c$,
and the asymptote at $|B_1-B_2|>B_c$ is
\begin{equation} 
\frac{\langle\tilde{G}_{max}(B_1)\tilde{G}_{max}(B_2)\rangle}
{\langle\tilde{G}^2_{max}(B_1)\rangle}
\approx  \frac{1}{16\sqrt{2}A_2}\frac{B_c}{|B_1-B_2|}\;.
\end{equation}
The full expression for $\langle
\tilde{G}_{max}(B_1)\tilde{G}_{max}(B_2)\rangle$ is too cumbersome to
present here.

 {\em Conclusion ---} In this paper mesoscopic fluctuations
of Coulomb blockade in a double-dot structure are studied.
Electron tunneling between the dots leads to the formation of peak doublets in
the dependence of conductance on gate voltages $V_{g1}$, $V_{g2}$. We
find the variance of spacings within the doublets, see Eq.~(\ref{deltaN}). This
variance provides information on the mesoscopic fluctuations of the ground
state energy of a partially opened quantum dot, which is hard to measure
directly. The pattern of the conductance as a function of $V_{g1}$, $V_{g2}$ is
affected by an unusually small magnetic field given by Eq.~(\ref{Bc}).

The authors are grateful to L.P. Kouwenhoven for valuable discussions.
This work was supported by NSF Grant DMR-9731756.

\end{multicols}
\end{document}